\documentclass[sn-mathphys,Numbered]{sn-jnl}

\usepackage{graphicx}
\usepackage{booktabs}
\usepackage{array}
\usepackage{amsmath,amssymb}
\usepackage{siunitx}
\usepackage{microtype}
\usepackage{manyfoot}
\usepackage{placeins}
\usepackage{xcolor}
\usepackage{tabularx}
\usepackage{multirow}
\usepackage{hyperref}
\usepackage[nameinlink,noabbrev]{cleveref}
\usepackage{csquotes}
\usepackage[T1]{fontenc}
\usepackage[utf8]{inputenc}
\usepackage{lmodern}

\graphicspath{{figures/}}
\hypersetup{hidelinks}
\newcommand{\keyfinding}[1]{\begin{center}\fbox{\parbox{0.92\linewidth}{\small\textbf{Key finding.} #1}}\end{center}}

\begin{document}

\title{Quantum-Classical Link Prediction on Dark Web Hyperlink Graphs}

\author*[1,2,3]{\fnm{Vojtěch} \sur{Novák}}\email{vojtech.novak.st1@vsb.cz}

\author[1]{Radovan Rečka}

\author[1]{Ondřej Knebl}

\author[1]{Aktham Youssef}

\author[1,2,3]{\fnm{Ivan} \sur{Zelinka}}

\affil*[1]{\orgdiv{Department of Computer Science, Faculty of Electrical Engineering and Computer Science}, \orgname{VSB-Technical University of Ostrava}, \city{Ostrava}, \country{Czech Republic}}

\affil[2]{\orgname{IT4Innovations National Supercomputing Center}, \orgname{VSB-Technical University of Ostrava}, \city{Ostrava}, \postcode{708 00}, \country{Czech Republic}}

\affil[3]{\orgdiv{Department of Informatics and Statistics, Marine Research Institute}, \orgname{Klaipeda University}, \city{Klaipeda}, \country{Lithuania}}

\abstract{%
We analyze HTML content and hyperlinks from an offline Tor crawl containing 383 onion hosts and 2,067 directed cross-host links. The study combines semantic analysis with degree-controlled quantum--classical link prediction. Two directory groups emit 94.8\% of the observed host links. Directed preferential attachment reaches ROC-AUC 0.976 with random non-edges, but falls to $0.550\pm0.014$ under source-conditioned target-degree matching. With disjoint training and test pairs, multimodal logistic regression reaches $0.672\pm0.021$, GAE $0.899\pm0.016$, and VGAE $0.906\pm0.016$. A four-qubit variational quantum classifier (VQC) comparison tests two encodings across 20 disjoint-pair graph splits under exact and 1,024-shot simulation, with classical controls using the same inputs and training examples. Its best exact mean ROC-AUC is $0.549\pm0.023$, compared with $0.551\pm0.016$ for RBF SVC and $0.573\pm0.029$ for a 16-parameter MLP on the same inputs and training examples. These results support predictive signal in combined graph/content features and show why degree control, feature compression, and optimization conditions matter in quantum--classical comparisons.
}


\maketitle

\section{Introduction}

Tor onion services form a web ecosystem that is difficult to observe completely. Services appear and disappear, discovery depends on crawl seeds and traversal rules, and directories can have a large influence on the observed hyperlink graph. Previous studies have examined Tor connectivity, content categories, community structure, redirects, and page/site relationships \cite{sanchezrola2017onions,avarikioti2018structure,bernaschi2022onion,li2026darkweb}.

Earlier crawls show large differences between the linking patterns of Tor services. Griffith et al. found that most sites in their sample had no outgoing links to other sites~\cite{griffith2017darkweb}. Bernaschi et al. described a small strongly connected core in which directories link to many peripheral services~\cite{bernaschi2022onion}. Zabihimayvan et al. studied page topics together with hyperlink structure to characterize service categories and their connections~\cite{zabihimayvan2019ecosystem}. Al-Nabki et al. proposed ToRank to identify influential onion domains from hyperlink connectivity and compared it with PageRank, HITS, and Katz~\cite{alnabki2019torank}. Node embeddings and graph auto-encoders also provide methods for learning features used in link prediction~\cite{grover2016node2vec,kipf2016vgae}. We study whether graph and content features predict withheld hyperlinks when the evaluation reduces the advantage of using endpoint degree alone.

Our dataset contains HTML snapshots and hyperlinks but no verified malware, activity, or service labels. We use the content to describe the sampled services and the observed graph to define a self-supervised link-prediction task. The DUTA study illustrates supervised content classification when activity labels are available~\cite{alnabki2017duta}. Al Nabki et al. later combined textual, HTML, visual, and topological features for supervised ranking, using manually ranked Tor services for evaluation~\cite{alnabki2023ranking}. For our study, the observed hyperlinks supply the prediction targets.

Link prediction is useful for studying incomplete networks, but its evaluation depends strongly on how absent edges are sampled. Common heuristics and learned models can exploit endpoint degree when positive and negative examples have different degree distributions \cite{wu2022linkprediction,li2023heart,aiyappa2025degree}. This is especially relevant in crawled web graphs with hubs and directories. We therefore measure this bias directly and use degree-matched negatives for the main comparison.

Quantum machine learning (QML) models can use quantum feature maps to introduce nonlinear interactions between numerical inputs~\cite{havlicek2019quantum}. Parameterized circuits are trained with classical optimizers, allowing shallow models to be studied with limited qubit counts~\cite{cerezo2021vqa}. Schuld et al. showed how data-encoding gates determine the Fourier frequencies available to a broad class of circuit models~\cite{schuld2021encoding}. In our setting, graph and content features must be compressed to fit a small qubit register. Link prediction therefore gives a task on which to test how much useful signal a compact circuit retains after compression. Huang et al. showed that training data can make classical learners competitive on quantum-motivated tasks~\cite{huang2021power}, and Bowles et al. found that standard classical models generally performed better than the quantum classifiers in their small-task benchmarks~\cite{bowles2024benchmarking}. Studies of variational energy minimization also report changes in optimizer performance under finite-shot noise~\cite{novakReliable2025,novakNoisy2026}. We compare variational quantum classifiers (VQCs) with classical models using the same inputs and training examples, and examine the effects of encoding, optimization, and finite-shot measurement.

Related QML studies include a framework for moving from classical to hybrid learning~\cite{illesova2025classical}, a combined convolutional and quantum model for image classification~\cite{illesova2025complementarity}, and a clinical comparison of quantum and classical classifiers~\cite{novak2026quantum}. Earlier quantum link-prediction studies include PQCLP, which uses trainable quantum circuit classifiers~\cite{singh2024pqclp}; PQKELP, which combines graph embeddings with projected quantum kernels~\cite{kumar2025pqkelp}; and QGCNLP, which applies quantum circuits to features aggregated within graph-convolutional layers~\cite{singh2026qgcnlp}. Our comparison uses a directory-dominated Tor hyperlink graph with degree-matched negatives and disjoint training and test pairs. Classical controls use the same inputs and training data, with additional comparisons for parameter count and optimization budget. The circuit design accounts for the influence of feature order and the measurement observable~\cite{salmenpera2025feature}.

This study combines semantic analysis of the crawl with a degree-controlled link-prediction benchmark. We measure how negative sampling affects the scores of degree-based methods, evaluate classical graph and content models on withheld links, and compare compact VQCs with classical controls. The comparisons account for input representation, training data, parameter count, and optimization.

\section{Methods}

\subsection{Data and content analysis}

The data are an offline subset of a Tor crawl. During collection, the hyperlink map was traversed breadth-first from an HTTP-200 page. Edge direction was ignored when selecting the connected component, and traversal stopped after reaching 1,000 HTTP-200 pages. Our analysis induces the HTTP-200 subgraph and uses its largest weakly connected component, containing 998 pages, 13,515 directed page-level hyperlinks, and an offline HTML snapshot for every page.

Pages are collapsed by onion hostname for the main graph analysis. A directed host edge $u\rightarrow v$ is created when at least one page on host $u$ links to a page on host $v$. Within-host links are removed from this graph. The result contains 383 hosts and 2,067 distinct directed cross-host edges. Of the page-level links with identified endpoints, 74.3\% stay within the same host. The host graph has density 0.0141 and reciprocity 0.0174; 75.5\% of hosts have no observed cross-host outgoing edge.

\begin{table}[t]
\centering
\caption{Summary of the analyzed dataset. Cross-host page edges count page-level hyperlinks, whereas host edges count distinct directed host pairs.}
\label{tab:data}
\begin{tabular}{lr}
\toprule
Quantity & Value \\
\midrule
HTTP-200 pages & 1,000 \\
Largest HTTP-200 weak component & 998 pages \\
Directed page edges in core & 13,515 \\
Onion hosts in core & 383 \\
Distinct directed cross-host edges & 2,067 \\
Host-graph density & 0.0141 \\
Host-graph reciprocity & 0.0174 \\
Zero cross-host out-degree hosts & 75.5\% \\
Zero cross-host in-degree hosts & 11.5\% \\
Parsed HTML snapshots & 998 \\
Pages containing exact token \texttt{porn} & 158 (15.8\%) \\
\bottomrule
\end{tabular}
\end{table}

HTML is parsed offline. Scripts and other non-visible elements are removed before text extraction. We use three text views: raw visible text; boilerplate-clean text with navigation, header, footer, and aside elements removed plus standard English stop words; and the same cleaned text with the token \texttt{porn} also removed. Host text is represented by TF--IDF over unigrams and bigrams and reduced to 32 dimensions with truncated SVD~\cite{manning2008ir}. Static counts such as forms, scripts, iframes, download-like links, Base64-like chunks, and selected JavaScript calls are used only as descriptive features because malware ground truth is unavailable.

Louvain communities \cite{blondel2008louvain} are computed on the undirected projection of the host graph. The observed partition has seven communities. For each text view, $k$-means~\cite{macqueen1967kmeans} is evaluated for $k=2,\ldots,12$. At $k=7$, agreement with Louvain communities is measured with normalized mutual information (NMI), adjusted mutual information (AMI)~\cite{vinh2010information}, and adjusted Rand index (ARI)~\cite{hubert1985partitions}.

For domain interpretation, we also build an exploratory semantic partition from boilerplate-clean text. Word- and character-level TF--IDF features are reduced to 64 dimensions with truncated SVD and clustered with $k$-means for $k=4,\ldots,12$. The highest cosine silhouette~\cite{rousseeuw1987silhouettes} in the tested range occurs at $k=12$ (0.237). We use the groups as a descriptive resolution because the best value occurs at the upper search boundary. Representative pages and top terms are inspected to assign short descriptive labels, and host links are aggregated into a directed cluster-to-cluster graph.

\subsection{Link-prediction evaluation}
\label{sec:lp-protocol}

The link-prediction experiment is transductive: all hosts are known and only edges are withheld. For each split, about 20\% of eligible undirected host dyads are selected for testing. A spanning tree is protected so that the training graph remains weakly connected. When a dyad is held out, every observed direction belonging to that dyad is removed, which prevents reverse-edge leakage.

In the primary benchmark, negative examples are unique non-edges of the complete observed host graph. Test negatives are reserved before training-negative sampling and excluded from the training pool, so the training and test pairs are disjoint. Degree, PageRank~\cite{manning2008ir}, adjacency embeddings, and all graph scores are computed from the training graph. Static text features remain available for all known hosts. The initial diagnostic benchmark uses random, topology-hard, and semantic-hard negatives. To reduce implicit degree bias in link prediction~\cite{aiyappa2025degree}, we then evaluate a directed degree-corrected sampler and use source-conditioned target-degree matching for the primary benchmark. For each test positive edge $(u,v)$, the negative keeps source $u$ and selects an absent target $w$ with training in-degree closest to that of $v$,
\begin{equation}
    \left|d_{\mathrm{in}}(w)-d_{\mathrm{in}}(v)\right|.
\end{equation}
Training negatives use the same matching rule where possible; when a source exhausts its unique non-edge targets, the remaining pairs are filled by directed degree-corrected sampling. This source fallback affects 230 training negatives per split (about 13.3\%); every test negative preserves its positive source. The primary results use 20 deterministic graph splits with seeds 42--61. Matching quality is measured with first Wasserstein distances between positive and negative log-degree distributions. Mean training distances are 0.339 for source out-degree and 0.195 for target in-degree, compared with 0.000 and 0.125 in testing.

Classical baselines include common neighbors, Jaccard coefficient, and Adamic--Adar~\cite{libennowell2007linkprediction}, together with directed preferential attachment,
\begin{equation}
    s_{\mathrm{PA}}(u,v)=d_{\mathrm{out}}(u)d_{\mathrm{in}}(v).
\end{equation}
A degree-only logistic model uses log source out-degree, log target in-degree, and their product. The graph representation combines truncated SVD of the training adjacency matrix and its transpose with degree features and PageRank. The content representation uses the 32-dimensional cleaned-text embedding. Their concatenation forms the multimodal representation. A directed pair $(u,v)$ is represented as
\begin{equation}
\phi(u,v)=\left[z_u,z_v,z_u\odot z_v,|z_u-z_v|\right].
\end{equation}
Class-balanced logistic regression is evaluated on graph, content, and multimodal features.

We use directed adaptations of the GAE and VGAE models of Kipf and Welling~\cite{kipf2016vgae} as relational baselines. Their node features combine training-graph degree/PageRank features with the static content representation. The encoder separates outgoing and incoming message passing,
\begin{equation}
 h' = W_{\mathrm{self}}h + W_{\mathrm{out}}\widetilde{A}_{\mathrm{out}}h + W_{\mathrm{in}}\widetilde{A}_{\mathrm{in}}h,
\end{equation}
where the outgoing and incoming operators separately row-normalize the training adjacency and its transpose. The hidden and latent dimensions are 48 and 24, respectively, followed by a directed decoder with separate source and destination projections. The VGAE adds a KL penalty of $10^{-3}$. Both models use Adam~\cite{kingma2015adam} for 80 epochs. Message passing uses only the training adjacency. Additional ranking and ablation fits use the same splits and 80-epoch protocol, with a gradient-checked NumPy implementation and fresh initialization. Reconstruction loss uses only explicit training pairs; loss-separation checks, ablations, and complete configurations are supplied in Online Resource 2.

ROC-AUC and average precision are reported over the 20 primary splits. We also report filtered all-target ranking for preferential attachment, multimodal logistic regression, and GAE/VGAE refits. For each held-out edge, the true target is ranked against all valid candidate hosts after filtering the source's other observed outgoing edges. Metrics are MRR, Hits@1, Hits@3, Hits@10, and median rank.

\subsection{Quantum comparison and resource accounting}
\label{sec:qml-methods}

Multimodal pair features are standardized using the complete training pool and reduced by principal component analysis (PCA)~\cite{jolliffe2002pca}; neither transformation uses held-out pairs. Components are scaled by their training maximum absolute value and clipped to $[-\pi,\pi]$. A classical sensitivity study evaluates PCA dimensions $d\in\{2,4,6,8,12,16\}$ across the 20 primary graph splits. Four and eight components retain about 17.7\% and 26.2\% of pair-feature variance, respectively.

A supplementary fixed-data optimizer study uses a one-repetition linear ZZ feature map~\cite{havlicek2019quantum} and RealAmplitudes ansatz implemented on eight qubits. Its single-qubit readout admits the exact reduced form $p_1=f(x_0,x_1,x_2;\theta_0,\theta_1,\theta_8)$, with zero-based indices, and the corresponding radial-basis-function support-vector classifier (RBF SVC)~\cite{cortes1995supportvector} control uses the same three PCA components. Ten matched runs use 256 training and 256 evaluation pairs from one seed-42 dataset, with common data and initialization per run. COBYLA~\cite{powell1994cobyla}, SPSA~\cite{spall1992spsa}, iSOMA/SOMA \cite{zelinka2004soma}, and jSO \cite{brest2017jso} minimize cross-entropy under a 300-function-evaluation (FE) limit in exact and 1,024-shot simulation; results are provided in Online Resource 1. CEC competition analyses provide further context for the use of evolutionary optimizers in variational quantum optimization~\cite{novakCEC2026}. This reduced-readout study supports only within-dataset optimizer comparisons, since it uses three effective inputs and parameters and 4.9\% of evaluation negatives recur in training on average.

The primary quantum comparison uses the first four components of the clean eight-component exports from all 20 primary splits. Each split has two balanced 256-pair training draws and independently initialized starts; every held-out pair is evaluated (668--680 pairs per split). Two feature maps are tested: product $R_y(x_q)$ encoding and the one-repetition linear ZZ map. Both use two layers of $R_z$ followed by $R_y$ on every qubit, with one linear CNOT chain between the layers. All-bit parity gives $p_1=(1-\langle Z_0Z_1Z_2Z_3\rangle)/2$. Gate-level checks agree with an independent dense-matrix implementation to below $1.4\times10^{-15}$; perturbation and Jacobian checks confirm all four inputs and all 16 parameters can affect the output. Including encoding, the circuits have 20 single-qubit gates and three CNOTs ($R_y$), or 27 and nine (ZZ).

COBYLA and SPSA minimize cross-entropy under a 200-FE limit, using the same training draw and initial angle vector for every encoding, optimizer, and simulator setting within a run. SPSA calibration calls count toward the budget. Batched NumPy statevector simulation supplies exact probabilities; finite-shot probabilities are binomial draws from the parity marginal with 1,024 shots per example. This models measurement sampling without gate or hardware noise. Similar exact and finite-shot comparisons have been used to assess numerical optimization for the Variational Hamiltonian Ansatz~\cite{illesovaVHA2025}. The full design comprises 320 VQC fits. We average the two starts within each graph split, then report means and standard deviations across the 20 splits as descriptive holdout variability.

Classical controls use exactly the same four inputs, preprocessing, training draws, and test pairs. The QMetric case study also compares quantum and classical models using the same PCA-reduced inputs~\cite{illesovaQMetric2025}. A parameter-matched tanh MLP has architecture $4\rightarrow3\rightarrow1$, no hidden biases, and an output bias: $4\times3+3+1=16$ trainable parameters. It is trained with the same 200-FE COBYLA and SPSA budgets and separately with gradient-based L-BFGS~\cite{liu1989lbfgs}. Logistic regression, RBF SVC, and a conventional 16-hidden-unit MLP (97 parameters) provide standard controls; a three-hidden-unit MLP with biases (19 parameters) is included in Online Resource 1. Equal parameter counts provide a capacity control; model expressivity and training cost can still differ. We retain per-fit weights, predictions, objective histories, and logical circuit/shot counts.

\section{Results}

\paragraph{Observed graph and semantic structure.}
Most page links remain inside a host: 9,947 are internal and 3,442 cross host boundaries. Host aggregation therefore removes a large amount of site navigation and leaves a sparse, strongly directional graph of 383 hosts and 2,067 cross-host links.

The content and graph partitions show weak agreement. At seven clusters, boilerplate cleaning raises NMI from 0.126 to 0.175 and AMI from 0.096 to 0.145, while ARI stays close to zero (\Cref{tab:agreement}). Removing \texttt{porn} after boilerplate cleaning leaves these agreement scores unchanged.

\begin{table}[t]
\centering
\caption{Agreement between seven content clusters and seven Louvain host communities.}
\label{tab:agreement}
\begin{tabular}{lrrr}
\toprule
Text view & NMI & AMI & ARI \\
\midrule
Raw visible & 0.126 & 0.096 & -0.008 \\
Boilerplate clean & 0.175 & 0.145 & $\approx 0.000$ \\
Boilerplate + \texttt{porn} filtered & 0.175 & 0.145 & $\approx 0.000$ \\
\bottomrule
\end{tabular}
\end{table}

The exploratory semantic partition provides a clearer view of the sampled content. Representative pages indicate groups related to drug storefronts, OnionMail, Tor search and navigation directories, carding-related storefronts, Russian-language darknet portals, and French-language local or alternative media (\Cref{tab:semantic-groups}). These labels summarize the observed pages; supervised activity labels were unavailable.

The directory groups have a distinct network role. The 47-host Tor search/link-directory group keeps only 6.4\% of its outgoing hyperlink weight inside its own group, and the smaller navigation-directory group keeps none. Together, these two groups emit 1,959 of 2,067 distinct directed host links (94.8\%) and point to several other semantic groups. Several specialized service/content groups have no observed outgoing cross-host weight in this crawl. The 12-group semantic partition also has weak agreement with Louvain communities (NMI 0.186, AMI 0.146, ARI 0.045).

\begin{table}[t]
\centering
\caption{Selected interpretable groups from the exploratory semantic partition. Labels summarize representative pages and top terms and are not ground-truth activity classes. A dash denotes no observed outgoing links, for which the internal fraction is undefined.}
\label{tab:semantic-groups}
\begin{tabular}{lrr}
\toprule
Descriptive group & Hosts & Internal outgoing weight \\
\midrule
French-language local/alternative media & 19 & --- \\
Russian-language darknet portals/marketplaces & 14 & 26.7\% \\
Drug-related storefronts & 13 & --- \\
OnionMail / encrypted mail & 22 & --- \\
Tor navigation directories & 8 & 0.0\% \\
Tor search/link directories & 47 & 6.4\% \\
Carding-related storefronts & 7 & 20.7\% \\
\bottomrule
\end{tabular}
\end{table}

The exact token \texttt{porn} occurs in 158 of 998 parsed pages (15.8\%). Under semantic-hard link evaluation, raw visible text reaches ROC-AUC $0.824\pm0.032$. Boilerplate cleaning and standard stop-word removal increase it to $0.874\pm0.046$. Adding \texttt{porn} to the stop-word list changes the result slightly to $0.877\pm0.046$. Most of the improvement therefore comes from removing repeated navigation and template text.

\paragraph{Degree bias in sampled link prediction.}
The initial benchmark gives very high scores to directed preferential attachment: $0.978\pm0.005$ with random negatives, $0.985\pm0.005$ with topology-hard negatives, and $0.989\pm0.006$ with semantic-hard negatives (\Cref{tab:initial-bias}). Learned models also score highly in these settings.

\begin{table*}[t]
\centering
\caption{Initial diagnostic benchmark. Values are mean ROC-AUC $\pm$ standard deviation across five repeated splits. These regimes are reported as a bias diagnostic; the primary benchmark uses degree-matched negatives.}
\label{tab:initial-bias}
\begin{tabular}{lccc}
\toprule
Model & Random & Topology-hard & Semantic-hard \\
\midrule
Directed preferential attachment & $0.978\pm0.005$ & $0.985\pm0.005$ & $0.989\pm0.006$ \\
Graph logistic regression & $0.969\pm0.006$ & $0.982\pm0.004$ & $0.930\pm0.035$ \\
Multimodal logistic regression & $0.967\pm0.007$ & $0.980\pm0.003$ & $0.914\pm0.032$ \\
Raw-content logistic regression & $0.961\pm0.004$ & $0.981\pm0.007$ & $0.824\pm0.032$ \\
Boilerplate-clean content & $0.961\pm0.006$ & $0.980\pm0.007$ & $0.874\pm0.046$ \\
Boilerplate + \texttt{porn} filtered & $0.961\pm0.005$ & $0.980\pm0.007$ & $0.877\pm0.046$ \\
\bottomrule
\end{tabular}
\end{table*}

\begin{figure}[t]
\centering
\includegraphics[width=0.82\linewidth]{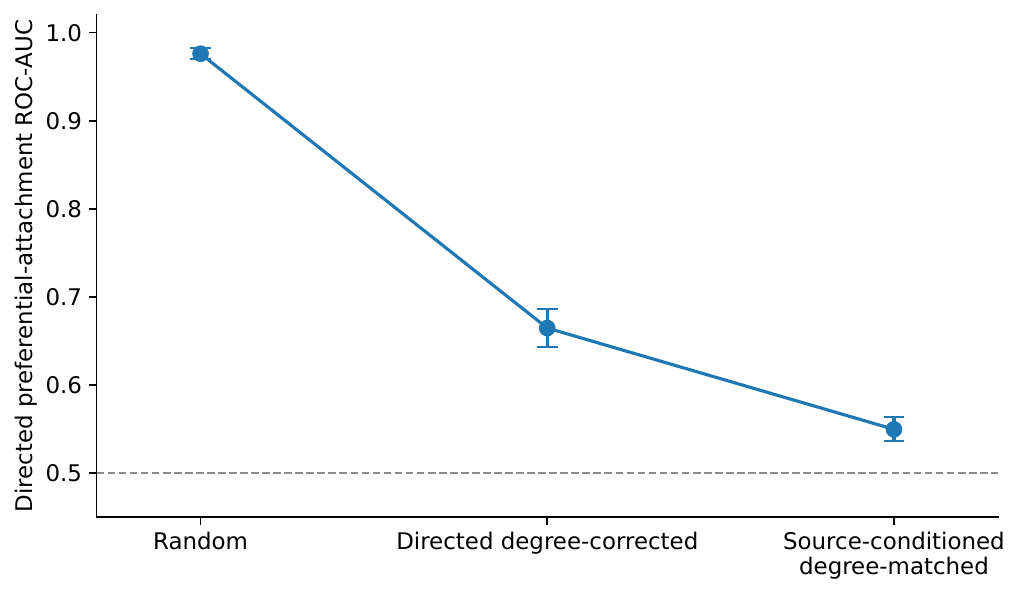}
\caption{Directed preferential-attachment ROC-AUC under increasingly degree-controlled negative sampling. Error bars show standard deviation over 20 splits.}
\label{fig:degree-audit}
\end{figure}

The 20-split audit shows that negative construction explains much of this performance (\Cref{tab:degree-audit,fig:degree-audit}). Preferential-attachment ROC-AUC falls from $0.9761\pm0.0064$ with random non-edges to $0.6648\pm0.0218$ with directed degree correction and $0.5498\pm0.0136$ with source-conditioned target-degree matching. The source out-degree Wasserstein distance falls from 3.146 to 0, and the target in-degree distance from 0.951 to 0.125.

\begin{table}[t]
\centering
\caption{Degree-bias audit across 20 held-out splits. $W_1$ denotes first Wasserstein distance between positive and negative log-degree distributions.}
\label{tab:degree-audit}
\begin{tabular}{lrrr}
\toprule
Negative sampler & PA ROC-AUC & $W_1$ source out & $W_1$ target in \\
\midrule
Random & $0.9761\pm0.0064$ & 3.146 & 0.951 \\
Directed degree-corrected & $0.6648\pm0.0218$ & 0.411 & 0.241 \\
Source-conditioned degree-matched & $0.5498\pm0.0136$ & 0.000 & 0.125 \\
\bottomrule
\end{tabular}
\end{table}

\keyfinding{Random non-edges make endpoint degree highly predictive in this crawl. Matching the source and target degree distributions reduces directed preferential attachment from $0.976$ to $0.550$ ROC-AUC.}

\paragraph{Models under degree control.}
Under source-conditioned target-degree matching, multimodal logistic regression reaches $0.6717\pm0.0212$ ROC-AUC, compared with $0.6253\pm0.0271$ for graph-only and $0.6327\pm0.0137$ for content-only logistic regression. Preferential attachment and the degree-only logistic model remain near 0.55. The directed graph autoencoders attain the highest mean scores: GAE reaches $0.8986\pm0.0158$ and VGAE $0.9058\pm0.0159$ across the same 20 splits (\Cref{tab:primary}). These results show predictive signal in the combined graph/content features under this test distribution.

\begin{table*}[t]
\centering
\caption{Primary source-conditioned target-degree-matched benchmark with disjoint training and test pairs. Values are mean $\pm$ standard deviation across 20 graph splits.}
\label{tab:primary}
\begin{tabular}{lcc}
\toprule
Model & ROC-AUC & Average precision \\
\midrule
Directed multimodal VGAE & $0.9058\pm0.0159$ & $0.8778\pm0.0230$ \\
Directed multimodal GAE & $0.8986\pm0.0158$ & $0.8673\pm0.0280$ \\
Multimodal logistic regression & $0.6717\pm0.0212$ & $0.6458\pm0.0181$ \\
8-D RBF SVC & $0.6429\pm0.0176$ & $0.6205\pm0.0174$ \\
Content logistic regression & $0.6327\pm0.0137$ & $0.6298\pm0.0129$ \\
Graph logistic regression & $0.6253\pm0.0271$ & $0.6130\pm0.0223$ \\
Directed preferential attachment & $0.5498\pm0.0136$ & $0.5681\pm0.0143$ \\
Degree-only logistic regression & $0.5476\pm0.0134$ & $0.5684\pm0.0148$ \\
Adamic--Adar & $0.5453\pm0.0197$ & $0.4995\pm0.0114$ \\
Common neighbors & $0.5344\pm0.0184$ & $0.5033\pm0.0104$ \\
Jaccard & $0.5329\pm0.0178$ & $0.5211\pm0.0144$ \\
8-D MLP & $0.5145\pm0.0085$ & $0.5409\pm0.0215$ \\
8-D logistic regression & $0.5138\pm0.0084$ & $0.5532\pm0.0079$ \\
\bottomrule
\end{tabular}
\end{table*}

\begin{figure}[t]
\centering
\includegraphics[width=0.92\linewidth]{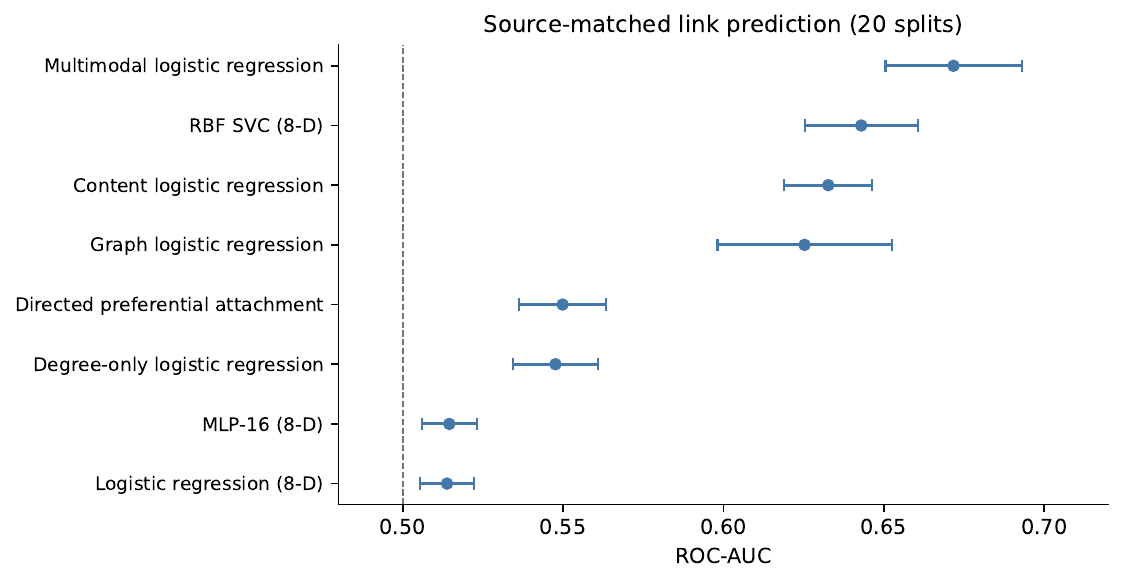}
\caption{Selected flat-feature models under the primary source-conditioned degree-matched protocol. Error bars show standard deviation over 20 graph splits.}
\label{fig:primary}
\end{figure}

The PCA sensitivity experiment measures compression of the pair representation under the same primary protocol. Mean explained variance rises from 12.1\% at two components to 26.2\% at eight and 35.9\% at 16. RBF SVC ROC-AUC rises from $0.563\pm0.019$ at two dimensions to $0.643\pm0.018$ at eight and $0.691\pm0.021$ at 16. Logistic regression stays near chance, while the small MLP remains between about 0.51 and 0.52.

\begin{table}[t]
\centering
\caption{Sensitivity of classical models to PCA dimension under the primary protocol. Explained variance and ROC-AUC are means over 20 graph splits; AUC values are reported as mean $\pm$ standard deviation.}
\label{tab:compression}
\begin{tabular}{rcccc}
\toprule
PCA dim. & Explained var. & Logistic regression & MLP-16 & RBF SVC \\
\midrule
2 & 0.121 & $0.515\pm0.007$ & $0.512\pm0.007$ & $0.563\pm0.019$ \\
4 & 0.177 & $0.511\pm0.006$ & $0.512\pm0.010$ & $0.620\pm0.018$ \\
6 & 0.223 & $0.515\pm0.008$ & $0.521\pm0.018$ & $0.636\pm0.016$ \\
8 & 0.262 & $0.514\pm0.008$ & $0.514\pm0.008$ & $0.643\pm0.018$ \\
12 & 0.315 & $0.511\pm0.009$ & $0.524\pm0.024$ & $0.667\pm0.022$ \\
16 & 0.359 & $0.511\pm0.010$ & $0.524\pm0.011$ & $0.691\pm0.021$ \\
\bottomrule
\end{tabular}
\end{table}

\begin{figure}[t]
\centering
\includegraphics[width=0.82\linewidth]{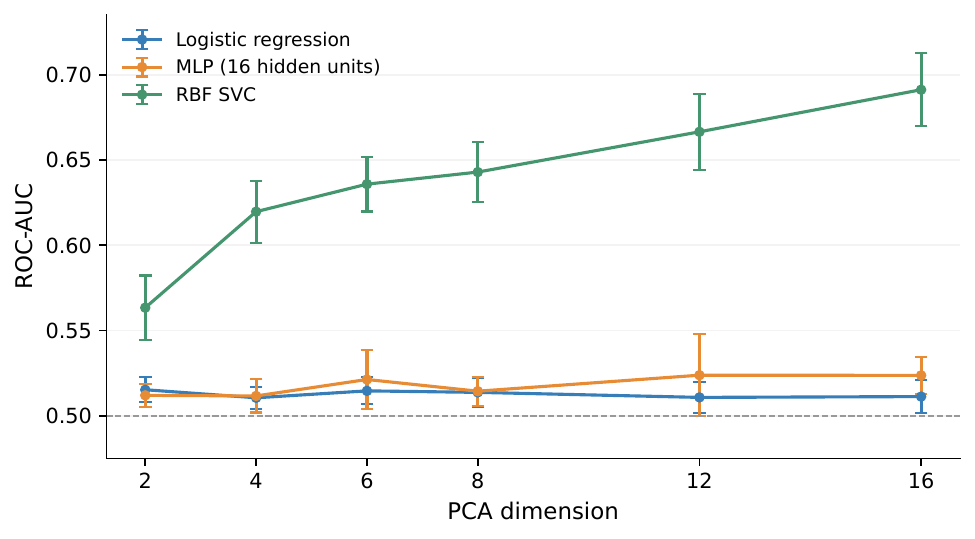}
\caption{Classical ROC-AUC as the PCA dimension is varied. Error bars show standard deviation over 20 graph splits.}
\label{fig:compression}
\end{figure}

Filtered all-target ranking gives a complementary view of degree. Preferential attachment has MRR $0.1645\pm0.0148$ and Hits@10 0.367, while multimodal logistic regression has MRR $0.0811\pm0.0135$ and Hits@10 0.218. Preferential attachment also has higher Hits@1 (0.070 versus 0.012). Thus the multimodal logistic model's higher sampled-pair AUC does not translate into better all-target ranking. The GAE/VGAE refits reach MRR $0.201\pm0.038$ and $0.214\pm0.039$, respectively, with Hits@10 of 0.504 and 0.528 (\Cref{tab:ranking}); the two tasks use different candidate distributions.

\begin{table}[t]
\centering
\caption{Query-weighted filtered all-target ranking, averaged over 20 graph splits; GAE/VGAE are additional refits on the same splits.}
\label{tab:ranking}
\begin{tabular}{lccccc}
\toprule
Model & MRR & H@1 & H@3 & H@10 & Median rank \\
\midrule
Preferential attachment & $0.1645\pm0.0148$ & 0.070 & 0.143 & 0.367 & 16.3 \\
Multimodal logistic regression & $0.0811\pm0.0135$ & 0.012 & 0.058 & 0.218 & 30.0 \\
GAE & $0.2013\pm0.0382$ & 0.072 & 0.216 & 0.504 & 10.5 \\
VGAE & $0.2137\pm0.0394$ & 0.085 & 0.222 & 0.528 & 9.7 \\
\bottomrule
\end{tabular}
\end{table}

\begin{figure}[t]
\centering
\includegraphics[width=0.82\linewidth]{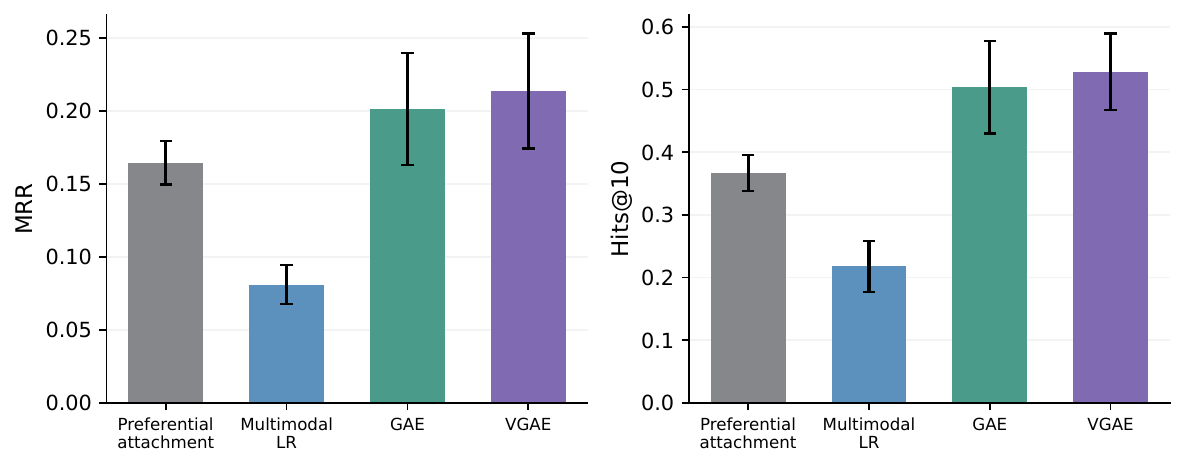}
\caption{Filtered ranking metrics for classical baselines and GAE/VGAE refits. Error bars show standard deviation over 20 graph splits.}
\label{fig:ranking}
\end{figure}

\paragraph{Quantum comparison.}
The four-qubit ZZ encoding attains the higher mean ROC-AUC in all four optimizer/simulator combinations (\Cref{tab:qml}). Its best exact mean is $0.549\pm0.023$ with COBYLA, compared with $0.528\pm0.010$ for the cheaper $R_y$ encoding under the same optimizer. With 1,024 shots, the ZZ means are $0.531\pm0.014$ (COBYLA) and $0.538\pm0.023$ (SPSA). Mean exact/shot FE are 200/174.6 for ZZ--COBYLA and 200/176.5 for $R_y$--COBYLA; SPSA uses 200 in both settings. A full-budget shot-based fit corresponds to $5.24\times10^7$ logical training shots.

On the same clean four-component inputs and 256 training examples, RBF SVC reaches $0.551\pm0.016$ ROC-AUC, the 16-parameter MLP trained by L-BFGS $0.573\pm0.029$, and the 97-parameter MLP $0.589\pm0.028$. With the same 200-FE COBYLA/SPSA budgets, the 16-parameter MLP instead reaches $0.523\pm0.013$ and $0.513\pm0.010$. The comparison therefore depends on optimization as well as parameter count. Complete AP, calibration, convergence, and resource results are provided in Online Resource 1.

\begin{table}[t]
\centering
\caption{Four-component comparison with disjoint training and test pairs. All models use the same balanced 256-pair training draws and full test pools. Values are mean $\pm$ standard deviation across 20 graph splits after averaging two starts per split. Classical scores are reported once. RBF SVC is parameterized by its support vectors.}
\label{tab:qml}
\begin{tabular*}{\linewidth}{@{\extracolsep{\fill}} lrcc @{}}
\toprule
Model / optimizer & Parameters & Exact ROC-AUC & 1,024-shot ROC-AUC \\
\midrule
VQC ($R_y$) + COBYLA & 16 & $0.528\pm0.010$ & $0.524\pm0.019$ \\
VQC ($R_y$) + SPSA & 16 & $0.520\pm0.027$ & $0.529\pm0.016$ \\
VQC (ZZ) + COBYLA & 16 & $0.549\pm0.023$ & $0.531\pm0.014$ \\
VQC (ZZ) + SPSA & 16 & $0.542\pm0.019$ & $0.538\pm0.023$ \\
\midrule
MLP + COBYLA & 16 & $0.523\pm0.013$ & --- \\
MLP + SPSA & 16 & $0.513\pm0.010$ & --- \\
MLP + L-BFGS & 16 & $0.573\pm0.029$ & --- \\
MLP (16 hidden units) + L-BFGS & 97 & $0.589\pm0.028$ & --- \\
RBF SVC & --- & $0.551\pm0.016$ & --- \\
Logistic regression & 5 & $0.510\pm0.007$ & --- \\
\bottomrule
\end{tabular*}
\end{table}

\begin{figure}[t]
\centering
\includegraphics[width=0.95\linewidth]{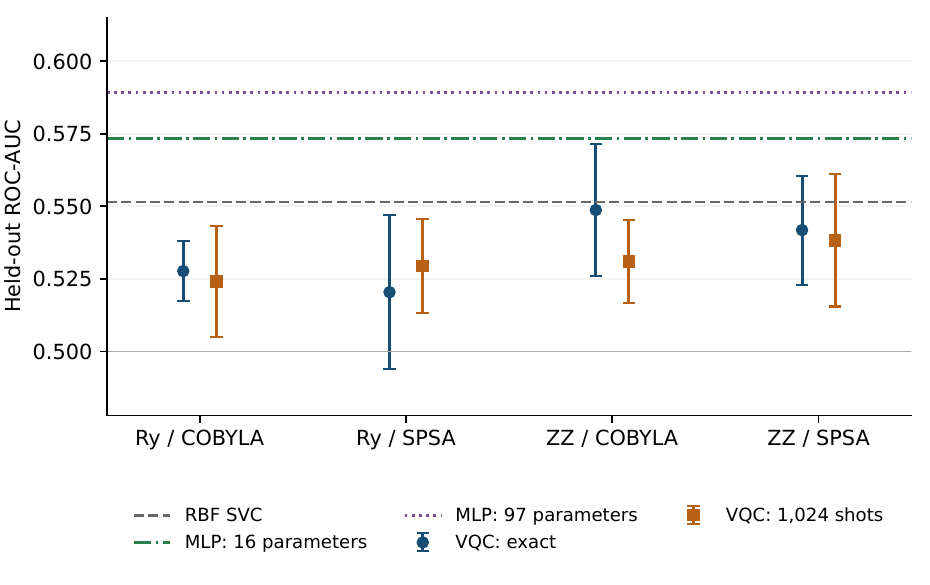}
\caption{Clean four-qubit comparison. Error bars show standard deviation across 20 graph splits, with two starts averaged within each split. The classical reference lines use the same inputs and training draws.}
\label{fig:qml}
\end{figure}

\keyfinding{With all four inputs and 16 parameters active, the best exact VQC mean AUC is 0.549, close to RBF SVC at 0.551. A 16-parameter MLP reaches 0.573 with gradient-based training. Matching inputs, parameter count, and optimization answers distinct comparison questions.}

\section{Discussion}

\paragraph{Graph structure and domain interpretation.}
The degree audit is central to the link-prediction results. Random and conventional hard negatives produce a large degree mismatch between positive and negative pairs, which makes preferential attachment appear nearly perfect. Source-conditioned target-degree matching reduces this shortcut in the test pairs. The multimodal GAE/VGAE models retain predictive signal with disjoint training and test pairs under this control. Ablations show that capacity-matched encoders without message passing reproduce sampled-pair AUC (about 0.900), but have lower MRR (0.157/0.152) than GAE/VGAE refits. Content-only GAE reaches 0.907 AUC, versus 0.769 with structure-only features; VGAE without its KL penalty reaches 0.899. Thus high sampled-pair AUC does not alone establish a message-passing or variational benefit. A four-endpoint-degree control reaches 0.574 AUC (Online Resource 2).

The semantic analysis helps explain what part of that structure represents. Directory and search services link outward across several content groups, while many specialized services behave mainly as destinations in the observed component. This gateway pattern also helps explain the weak agreement between semantic clusters and Louvain communities. Content and topology describe different aspects of the crawl, and models that combine them have access to both. This directional asymmetry is consistent with the sparse linking and directory-centered structure reported in earlier crawls~\cite{griffith2017darkweb,bernaschi2022onion}, while jointly considering content and hyperlinks follows the perspective of Zabihimayvan et al.~\cite{zabihimayvan2019ecosystem}. Our extension quantifies how the observed asymmetry affects negative-sampling bias and the interpretation of predictive performance.

Source-preserving training negatives change GAE/VGAE mean AUC by only $+0.001$ each; dropping fallback-aligned loss pairs lowers both by about 0.019. Outside directory sources, their AUC is 0.788/0.779, with only 6--26 positive test edges per split. All-target ranking is query-weighted; under equal source weighting, GAE/VGAE MRR is 0.145/0.152 versus 0.147 for preferential attachment. Sampled-pair AUC and ranking measure different candidate distributions.

\paragraph{Classical and quantum comparison.}
Compression is a material restriction: RBF SVC reaches about 0.691 ROC-AUC with 16 PCA components under the full primary training protocol, whereas four components retain only 17.7\% of pair-feature variance. A sample-size control raises four-component RBF AUC from 0.551 to 0.620 using the full training pool; with 256 examples, eight components yield only 0.553. The four-qubit circuits remain close to the matched RBF control and below the mean of the parameter-matched MLP trained by L-BFGS. Their advantage over that MLP under the same gradient-free budget does not establish a general quantum advantage; the stronger gradient-based control makes the optimization dependence explicit.

The encoding comparison identifies a modest benefit from ZZ features in the tested configurations, at the cost of six additional CNOTs. Exact ZZ--COBYLA fits reach mean training AUC 0.717 versus 0.549 held out. Under shot-based training, their exact held-out mean falls to 0.531, whereas resampling fixed exact-fit predictions gives AUC SD about 0.003. These diagnostics implicate limited supervised data and optimization under sampling (Online Resource 2). FE, circuit instances, and shots clarify computational resources; small-state batched simulation time does not estimate hardware execution time. The finite-shot results concern measurement sampling only. The encoding comparison is motivated by the role of data encoding in circuit expressivity~\cite{schuld2021encoding}. The need for strong classical controls also follows the benchmarking findings of Bowles et al.~\cite{bowles2024benchmarking}. Our study complements previous circuit-based, projected-kernel, and hybrid graph-convolutional approaches to quantum link prediction~\cite{singh2024pqclp,kumar2025pqkelp,singh2026qgcnlp} by examining graph sampling, feature compression, and optimization together in a Tor hyperlink task. Comparisons across these studies remain qualitative because the datasets, model families, and evaluation protocols differ. These experiments support a controlled application benchmark rather than a claim of quantum superiority.

\paragraph{Limitations.}
The crawl was deliberately built as a connected BFS component, so seed choice and traversal affect the observed degree and content distributions. One snapshot supports edge reconstruction among known hosts; temporal forecasting would require repeated crawl snapshots. Most hosts have zero observed cross-host out-degree, so source-conditioned evaluation mainly concerns link-emitting hosts. The GAE/VGAE setting is transductive: graph propagation uses the training graph while static content is available for all known hosts. The primary classical and four-qubit studies use 20 disjoint-pair splits. Repeated holdouts of one graph quantify split variation, not generalization across crawls. The quantum models use strongly compressed features, small training subsets, and shallow circuits without exhaustive hyperparameter search. Finally, the semantic groups are exploratory; some groups reflect topic, language, shared templates, or mirrors, and the descriptive labels are exploratory and do not provide verified activity classes or prevalence estimates. Static HTML indicators are descriptive features; malware ground truth is unavailable.

\section{Conclusion}

We analyzed 383 Tor onion hosts using their HTML content and 2,067 directed cross-host links. The content analysis finds recognizable service themes and shows that directory/search groups act as outward gateways across several groups. Content and topology have weak community agreement, which supports treating them as complementary views.

The main link-prediction result is the effect of degree control. Directed preferential attachment reaches about 0.976 ROC-AUC with random non-edges and about 0.550 after source-conditioned degree matching. Under disjoint-pair evaluation, multimodal logistic regression reaches about 0.672, GAE about 0.899, and VGAE about 0.906 ROC-AUC. These results show predictive signal in the combined graph/content features after controlling source out-degree and target in-degree in the test pairs.

The quantum comparison comprises 320 four-qubit fits evaluating two encodings under exact and finite-shot simulation across 20 disjoint-pair graph splits. The best exact VQC mean AUC is about 0.549, close to the matched RBF control at 0.551; a 16-parameter MLP trained with gradients reaches 0.573. These findings make the role of compression, measurement, and optimization explicit. Together with the degree audit, they provide a reproducible quantum--classical benchmark for the sampled hyperlink graph without establishing a general quantum advantage.

\section*{Acknowledgements}
This project has received funding from the Research Council of Lithuania (LMTLT), agreement No.: S-ITP-25-7. This research was also supported by research grants SGS No. SP2026/063 of VSB-Technical University of Ostrava, Czech Republic.

\section*{Data availability and reproducibility}

All simulations were conducted using a locally developed code, which is openly available for further examination and reproduction of results at \url{https://github.com/VojtechNovak/dark-web}.

\section*{Data and ethics statement}
The dataset may contain sensitive or illicitly advertised content. Results are reported at aggregate level and onion addresses are not published. Semantic labels are descriptive summaries of unsupervised groups, and static HTML indicators are not used as claims of malicious behavior.

\small
\bibliography{references}
\end{document}